\documentclass{llncs}
\usepackage{graphicx}
\usepackage{amsmath}
\usepackage [autostyle]{csquotes}
\usepackage{soul}

\usepackage[square,sort,comma,numbers,sectionbib]{natbib} 
\usepackage{url}

\begin{document}
\mainmatter              % start of a contribution
\title{\Large Death, Taxes, and Inequality }
\subtitle{\normalsize Can a Minimal Model Explain Real Economic Inequality?} 
\titlerunning{Death, Taxes, and Inequality}  % abbreviated title (for running head)
\author{John C. Stevenson }
%\orcidID{0000-0001-8518-9997}}

%
\authorrunning{JC Stevenson} % abbreviated author list (for running head)
%Craig Chambers, Kim B. Bruce, and Elisa Bertino}
%

\institute{Long Beach Institute, Long Beach, NY 11561\\
\email{jcs@alumni.caltech.edu}
}
%\orcidID{0000-0001-8518-9997}

%\\ WWW home page:\texttt{http://users/\homedir iekeland/web/welcome.html}
%}

\maketitle              % typeset the title of the contribution

\begin{abstract}
	
	Income inequality and redistribution policies are modeled with a minimal, endogenous model of a simple foraging economy.
	Significant income inequalities emerge from the model for populations of equally capable individuals presented with equal opportunities.
	%The model is scaled to match human lifespans and overall death rates. %simulates actual redistribution policies to
       %  Empirical and model data are fit to implied distributions providing necessary resolution for comparison. 
	Stochastic income distributions from the model are compared to empirical data from actual economies.
	%	These model explanations are compared to actual income distributions and actual effectiveness of redistribution policies. 
	 %The stochastic nature of the model allows more nuanced explanations. %as well as investigation of \textquote{what if} scenarios not possible with actual economic data. 
	The impacts of redistribution policies on total wealth, income distributions, and inequality are shown to be similar
	for the empirical data and the model.
	 %in the model and in the actual data 
	%appear similar while income taxes reduce inequality in actual data but not in the model. 
	%	The dangers of decreasing fertility highlight the usefulness of this stochastic model.
	These comparisons enable detailed determinations of population welfare beyond what is possible with total wealth and inequality metrics.
%	Estate taxes in the model appear quite effective in reducing inequality without reducing total wealth.
	
%	Stochastic population instability at both the high and low ends of infertility are considered.
	
	\keywords{ income inequality, income redistribution, ABM
	%,stochastic population dynamics
	}
\end{abstract}
\section{Introduction}
This research compares empirical measurements of income distributions in modern economies with income distributions 
that emerge from a minimal model of a foraging economy under various redistribution policies.
Income and wealth inequalities are subjects of interest to the general public \citep{marx2023kapital,piketty2022brief}, governments \citep{dabla2015causes,taghizadeh2020impact}, economists \citep{ravallion2014income,smith2001wealth}, sociologists \citep{wilkinson2009income}, epidemiologists \citep{tibber2022association,muntaner2020income}, and political scientists \citep{huijsmans2022income,engler2021threat}.  These inequalities are often attributed to a systemic lack of educational and employment opportunities \citep{hoffmann2020growing,dabla2015causes} and are also seen as results of politics and policies \citep{polacko2021causes,dabla2015causes}. Many government policies attempt to address these inequalities through redistribution of income by taxation \citep{causa2017income,taghizadeh2020impact}.
The complexity of the economy thwarts any clear assignment of causations and cures for inequality \citep{doran1999forecasts,sharma2014ever},
%Effects of policy changes are opaque at best and often unmeasured. The policy is often changed due to political processes before the policies' effects have had a chance to be manifest.  
resulting in many contradictory explanations \citep{muntaner2020income,davies2017wealth,trabandt2011laffer}.

Conversely, a minimal model of a system \citep{roughgarden}, in this case an agent-based, endogenous model of a simple foraging economy, provides repeatable, quantifiable, and stochastic explanations of inequality.  
Resource distributions of entire populations emerge based on simple behaviors of underlying agents and the landscape characteristics on which they forage.
Actual income distributions and redistribution policies are compared with distributions that emerge from the model employing similar redistribution policies and scaled to human lifespans.

%First, the baseline model is described and placed in context with the appropriate mathematical models of biology, ecology and genetics.  
%The scaling of the model to human lifespans is described, and implementations of various redistribution polices are detailed.
%Representative samples of empirical economic data are introduced and the need for implying income distributions is argued. 
%Shortcomings with the sampled data and with inequality metrics are discussed with emphasis on rich tail distributions. 
%A discussion of the correspondence of the model's resources to actual economy's income and wealth is provided. . 
 %Actual  income redistribution policies including estate taxes are selected for comparison with the model. 
% The income distributions for the empirical data and the model are compared and discussed under various resource redistributions. 
%These stochastic simulations of population dynamics isolate the effects of particular policies and allow exploration of novel explanations.

\section{Methods}

%\subsection{Introduction} 
A spatiotemporal, multi-agent-based model based on Epstein and Axtell's classic Sugarscape \citep{axtell,stevenson} is used to model a simple foraging economy \citep{stevenson2021population,stevensonEcon}. As a minimum model of a system \citep{roughgarden}, the model does not attempt to calibrate to an empirical objective function. Rather, a population of agents endogenously evolves under evolutionary selection pressures. 
%Some models in this category apply selection pressure exogenously, for example Iterated Prisoner's Dilemma contests \cite{axelrod,fogel,lindgren,lindgrenNordahl,miller,skyrms,skyrmsB}. Others apply the selection pressure endogenously within the simulation as a \textcquote{gause}{struggle to survive}, where more fit individuals reproduce and replace the less fit. Well-known examples of endogenous selection in a minimal model of a system are the aforementioned Sugarscape , Pepper and Smuts's alarm calling and feeding restraint model \cite{pepper}, and a demographic Prisoner's Dilemma study \cite{epsteinIP}. 
The foraging resources are evenly distributed across the landscape giving equal opportunity to all. The capabilities of each agent are identical.
%\footnote{The baseline model specifications  \citep{stevenson2023local}.}. 
%Surprisingly, even with these equal opportunities and capabilities, significant wealth inequalities emerge in both the dynamic and steady state regimes of the model \citep{stevenson2023local, stevensonEcon}.  

\hl{}For this simple model to provide explanations worthy of serious consideration, it must agree with standard models of discrete populations, stochastic genetics, ecology, and complex adaptive systems.
The dynamics that emerge from this simple underlying model have been shown to agree with discrete Hutchinson-Wright time delayed logistic growth model of mathematical biology and ecology \citep{murray,stevenson,hutch1,kot,stevensonEcon}, with standard Wright-Fisher class, discrete, stochastic, gene-frequency models of mathematical population genetics for finite populations, \citep{ewens,moran,cannings,stevenson}, and with modern coexistence theory for multiple species.\citep{chessMech,stevensonX}.  Dynamics of complex adaptive systems emerge with both intra-group and intergroup evolutionary optimizations \citep{wilsonCAS,stevensonX,stevensonGP}. \hl{}Conversely, almost all essential mechanisms of actual modern economies are not included with the hope that fundamental insights into complex economies may be gained by comparison with the simple model.
\hl{}While the model does also provide spatial (unmixed) distributions of income, the possible segregation of incomes levels and its effects on local sharing and commons is beyond the scope of this paper.

%Of particular interest are stable, oscillatory, and chaotic regimes determined by the intrinsic rate of growth.
  \begin{figure}
	\begin{center}		
	     \resizebox{\columnwidth}{!}{	
		\includegraphics[angle=-90,scale=1.0]{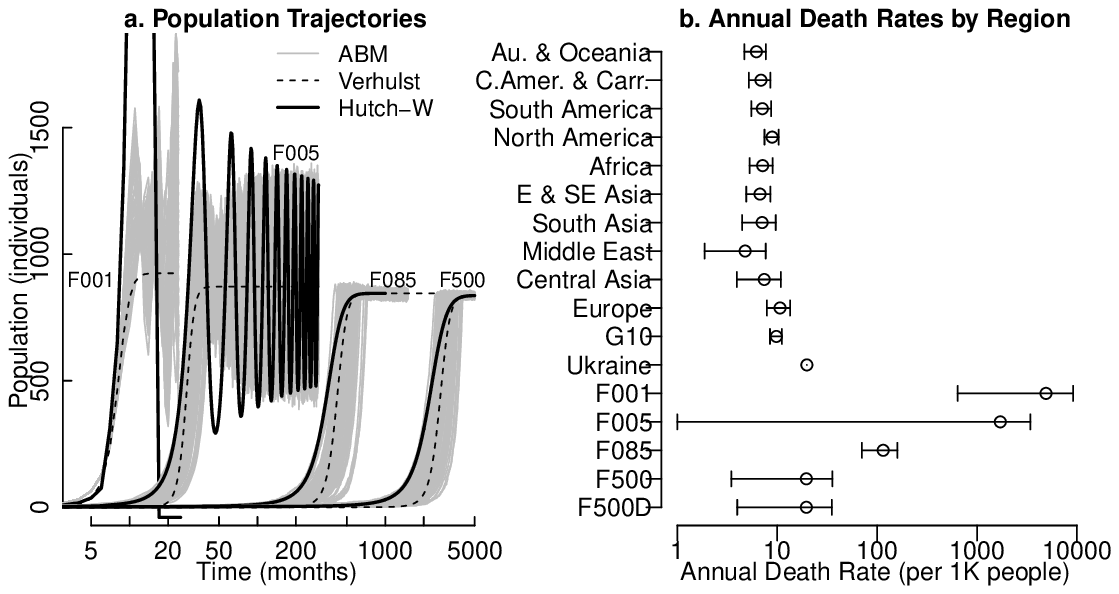}
		}
	\end{center}
	\caption{\small Population Trajectories and Global Death Rates. \scriptsize a)Various dynamic population regimes as functions of infertility. b) Annual Death Rates for geographic regions \citep{ciaDeathR} and for various fertilities in the model (only F500D has FL).}	
	\label{fig:popTandDeath}
\end{figure} 

\subsection{Configuring the Simulation to Actual Economies}
The intrinsic growth rate of this configuration is modulated by infertility.  Given birth cost and free space constraints are met, the probability of reproduction, $p_{f}$, is expressed as infertility $ f = \frac{1}{p_{f}}$ and labeled as F$p_{f}$ (e.g. F010 has a 10\% probability of reproduction). 
%Though the chaotic (F001), oscillatory (F005), and stable (F085) regimes are of great interest, they do not reflect the death rates of current human societies as shown by Figure \ref{fig:popTandDeath}b \citep{ciaDeathR}. 
Figure \ref{fig:popTandDeath}a highlights the population dynamics for various regimes. An increase to infertility F500 achieved death rates comparable with modern human societies \citep{ciaDeathR} as shown by Figure \ref{fig:popTandDeath}b .
%\begin{figure}
%	\begin{center}		
%	     \resizebox{\columnwidth}{!}{	
%		\includegraphics[angle=-90,scale=1.0]{mortCia2.eps}
%		}
%	\end{center}
%	\caption{\small Death Rates} 
%	\label{fig:deathRates}
%\end{figure}
%https://www.cia.gov/the-world-factbook/field/death-rate/country-comparison/}

Agents' ages were limited by scaling action cycles to human lifespans. Defining 10 action cycles as a year, agent deaths due to \textquote{natural causes} are generated by a finite lifetime (FL) heuristic with a flat probability of death between 60 and 100 years. The addition of FL had no measurable effect on death rates (Figure \ref{fig:popTandDeath}b F500D), though forty percent of the population was now dying of FL rather than starvation.  Two options for were added for inheritance, one bequeaths equally to all direct, surviving offspring, and the other imposes a 100\% estate tax.  

\subsection{Redistribution of Resources }

%To support comparison with modern economies, top down taxing was implemented, with both a monthly tax on the wealthy's income and an estate tax at death.
There are numerous approaches for redistribution of income and wealth, both from rich to poor and, surprisingly, poor to rich \citep{causa2017income,taghizadeh2020impact}. In empirical studies of inequality and taxation, a strong distinction is drawn between \textquote{income} and \textquote{wealth}  \citep{costa2019not,davies2017wealth}. \hl{}While a large body of work has been performed on income distributions and taxation, much less work has been done with wealth distributions due to the difficulty in defining wealth, measuring wealth, and taxing wealth \citep{costa2019not,drometer2018wealth,davies2017wealth}. \hl{}Single point metrics of inequality based on wealth or income can be surprisingly contradictory. (Sweden in 2022 had a Gini Coefficient (GC)  based on income of 0.298 which was 137th globally but had a wealth-based GC of 0.881 which was top twelve globally \citep{wbGini,csGWD}). \hl{}This research's objective is to compare the welfare of whole populations rather than abstract inequality measures. In fact, Domhoff \citep{domhoff2011wealth} \hl{}reports the financial wealth of the lower 80\% of the United States population in 2012 possessed only 4.7\% of the total wealth, making wealth without wealth taxation essentially immaterial for the welfare of the vast majority of the population.

Extensive studies of income redistribution across OECD countries gives the changes of inequality due to income redistributions as measured by GC \citep{causa2017income}.  These studies show significant percentage reductions in GC through income redistribution. The mean OECD GC reduction was 29\% due to redistribution. Use of the change of GC among various countries rather than the actual GC is in recognition of the inadequacy of GC for comparisons of sampled populations of different sizes \citep{fontanari2018gini}.
%Empirical data on wealth and income distributions show substantial differences in respective inequality. For example, the income inequality for Sweden measured by the Gini Coefficient is 29.8\% ranking 137th globally \citep{wbGini} while the wealth inequality measure is at 88.1\%, putting it the top twelve globally\citep{csGWD}
Estate taxes do not generate meaningful tax revenues for redistribution  %enjoy some academic praise \citep{caron2012occupy,drometer2018wealth,hoover1927economic,aaron1992reassessing,bird2013death}, they 
\citep{bird2013death,caron2012occupy}, perhaps because they are very difficult to administer and easy to evade. On average, among OECD countries, estate and gift taxes make up 0.1\% of GDP, while total tax revenues account for 34.3\%, %Initially, a local sharing tax on neighbors as an income redistribution method was implemented in the spirit of simple, bottom up agent behaviors \citep{stevenson2023local}. 
\hl{}While these particular estate taxes may impact single point metrics of inequality, due to the insignificant amount of revenue they generate in modern economies, they have no material effect on overall welfare.

%The question then arises, should the surpluses acquired by agents by foraging be considered wealth or income? Though foraging is income, storage over time suggests wealth, as does inheritance. The distinction that wealth generates additional income and surpluses do not supports treating surplus as income. Adjustments are easily made for gross or disposable income since the metabolism cost each action cycle is uniformly 3 resources.
%\subsubsection{Surplus Taxation and Redistribution}

A top-down redistribution of resources through a monthly tax on the richest individuals was implemented for comparison with actual redistribution economies. \hl{}All the revenues from this monthly income tax are fully redistributed to the poor. A percentage of the total surplus is defined (tax bracket) and those richest agents in that surplus tax bracket contribute one resource each month to a global pool. %and all the individual agents whose surplus is within this bracket 
This pool is distributed to the poorest agents, with preference to the older agents.

An option to allow inheritance was implemented by transferring the surplus of a dying agent equally to all surviving children (but not grandchildren). If there are no surviving children, the surplus is lost.  A 100\% estate tax option was also implemented in which surplus of any agent dying is lost and not redistributed. \hl{}The 100\% estate tax is the only tax where revenues are not fully redistributed. 

\section{Implied Distributions}      
%\subsection{Introduction}

%Economic data on distributions of income and wealth, and the effects of redistribution on these distributions, have been developed since at least as the early 20th century \citep{hoover1927economic}. Recent empirical studies show mixed success with reducing inequality with some redistributions actually transferring wealth from poor to rich  \citep{causa2017income} . .

%( https://data.worldbank.org/indicator/SI.POV.GINI/) 
%(Credit Swiss .  Wealth DataBook 2021 https://www.credit-suisse.com/media/assets/corporate/docs/about-us/research/publications/global-wealth-databook-2022.pdf). 

\begin{figure}
	\begin{center}		
	     \resizebox{\columnwidth}{!}{	
		\includegraphics[angle=-90,scale=1.0]{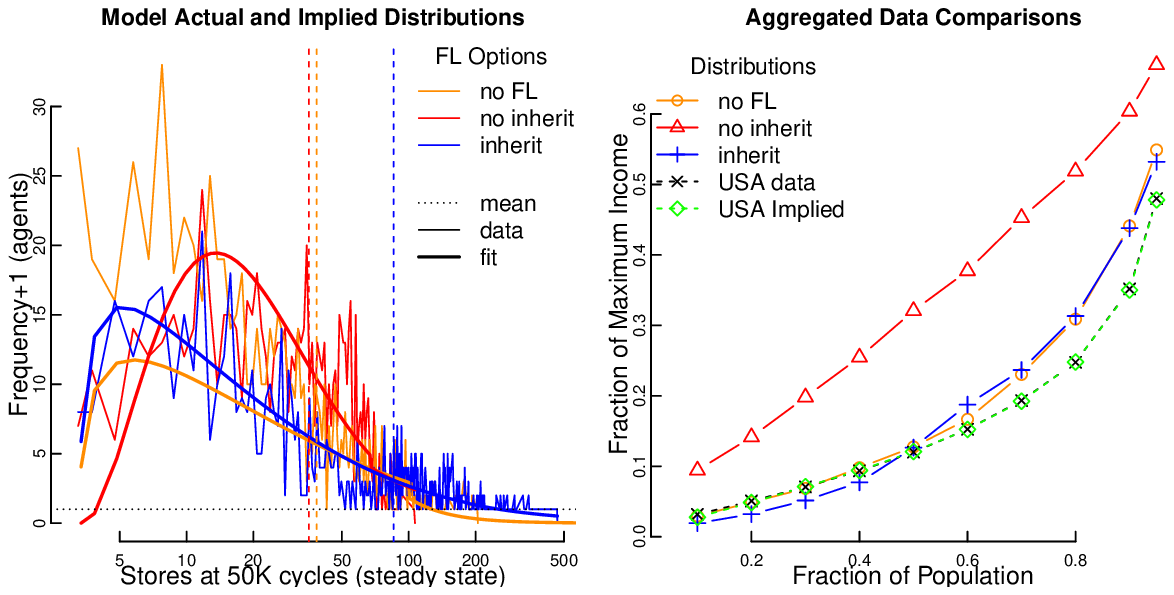}
		}
	\end{center}
	\caption{
		 Actual, Implied, and Aggregated Distributions.  \scriptsize a) Model's Actual and Implied Distributions. Three no-sharing configurations with and without FL and estate taxes, showing the actual data and the implied hybrid distributions. b) \hl{}Aggregated Data Comparisons. The actual USA aggregated data shown with the same aggregation of its own implied distribution and three simulated actual distributions.} 
	\label{fig:modelAndActual}
\end{figure}

Income distributions for actual economies are almost never published on an individual level, instead low resolution aggregations are provided. Conversely, the model generates actual incomes for all individuals.  In order to compare these very different representations, both aggregated (Figure \ref{fig:modelAndActual}b) and implied income distributions are generated. An implied approach using a lognormal distribution with a Pareto distribution for the upper tail, while not the most advanced mathematical approach, has been demonstrated to be sufficiently accurate \citep{druagulescu2001exponential,charpentier2022pareto} and provides insight into potential distributional differences in the upper tail relative to the rest of the population and between empirical and simulated distributions.

\begin{table}[h!]
	\begin{center}
	\begin{tabular}{|c|c|c|c|c|c|c|c|c|c|c|c|c|}
		\hline
		Economy& $N_{s}$ & $GC_{a}$& $\sigma_{L}$ & $\mu_{L}$ & $rms_{L}$ & $GC_{L}$ & $\sigma_{H}$ & $\mu_{H}$ & $\alpha$  & $rms_{H}$  & $GC_{H}$ \\
		\hline
		UK Income& 80& 0.466 & 0.52 & 10.5 & 937 & 0.475 & - & - &-&- &-\\
		UK decile& 10&0.370 & 2.5 & -0.15 & 2.6K & 0.416 & 1.36 & -0.16& 1.5  &260.1 &0.373 \\
		USA & 10&0.414& 1.9 & -0.25 & 53K & 0.412& 1.2& -1.1& 1.7 & 3343 &0.411 \\
		no FL & 134 &0.364& 1.6 & -1.6 & 27.0&0.454& 1.7 & -1.45& 0.6 & 34.4 &0.459\\
		FL no inherit& 88&0.325& 1.0 & -1.1 & 35.8 &0.337& 1.1 & -1.1& 0.6 & 34.4 &0.346\\
		FL inherit & 246&0.386& 1.9 & -1.3 & 29.6&0.447 & 2.1& -1.1& 0.1& 29.2&0.448\\
%		FL inherit quad      & 1741& 0.435 & 3.5 &5.48&38.4&0.220& 3.55 &5.52&  0.1&38.5&0.271\\
%		FLInherit IR & 284 &0.424 & 2.1 & -1.25 & 25.81 &0.467&2.2&-1.1&  0.1& 25.79& 0.421\\
		\hline
	\end{tabular}
\hfill \break
	\end{center}
	\caption{Implied Distribution Parameters, Fit Errors, and Inequality Measures. (\scriptsize $N_{s}$ is the number of samples for each data set, $GC_{a}$, $GC_{L}$ and $GC_{H}$ are the Gini Coefficients for the empirical data, lognormal, and hybrid distributions respectively,   $\sigma_{L}^2$ is the variance and $\mu_{L}$ is the mean of the lognormal only distribution, and $\alpha$ is the Pareto shape factor and $x_{m}$ is the threshold for the Pareto regime, $rms_{L}$ and $rms_{H}$ are fit errors for lognormal and hybrid distributions, respectively.)}
	\label{tab:fits}
\end{table}

\begin{figure}
	\begin{center}		
	     \resizebox{\columnwidth}{!}{	
		\includegraphics[angle=-90,scale=1.0]{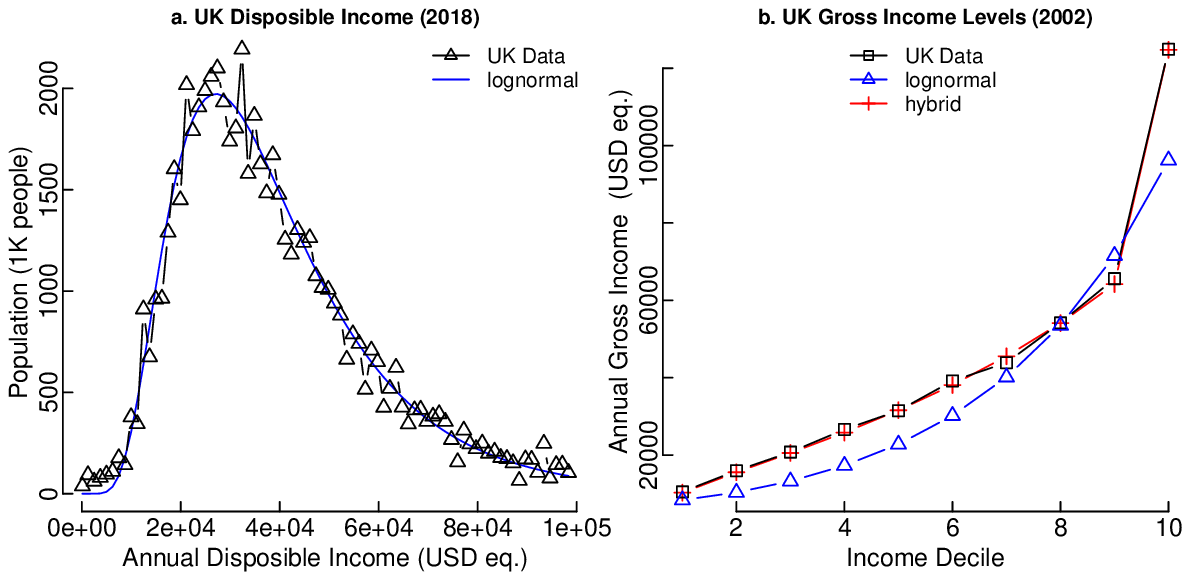}
		}
	\end{center}
	\caption{\small Implied Distributions for UK Data. \scriptsize a)The empirical data and lognormal fit for the 2018 UK data  \citep{uk2018}. b) The empirical decile data and hybrid and lognormal only fits for the UK 2002 data \citep{uk2003} highlighting the benefit of the Pareto fit to the rich tail. }
	\label{fig:ukFits}
\end{figure}

The first data set considered provides  mean disposable incomes for the UK in 2018 with  \pounds 1K  bins from \pounds 0 to  \pounds 79K (number of samples $N_{s}=80$) \citep{uk2018}. Therefore, \pounds 79K is the reported maximum disposable income. %Figure \ref{fig:logIndividual}a shows the solution space and RSS errors over the mu and sigma parameters and Figure \ref{fig:logIndividual}b the resultant best fit to the entire dataset. 
Figure \ref{fig:ukFits}a presents the empirical data and its lognormal fit which was stable and accurate for the entire range of incomes provided. (All fitted parameters for implied distributions and RMS errors are reported in Table \ref{tab:fits}.) While unexpected, this successful lognormal-only fit may be due to the reported maximum disposable income of only \pounds 79K, a likely truncation of higher earners.

%https://www.ons.gov.uk/peoplepopulationandcommunity/personalandhouseholdfinances/incomeandwealth/bulletins/householddisposableincomeandinequality/yearending2018

%\begin{figure}
%	\begin{center}		
%	     \resizebox{\columnwidth}{!}{	
%		\includegraphics[angle=-90,scale=1.0]{ukIncomeDfit.eps}
%		}
%	\end{center}
%	\caption{\small Lognormal  for UK Individual Income Distribution}
%	\label{fig:logIndividual}
%\end{figure}   
The next dataset, mean gross income by decile for the UK in 2002 \citep{uk2003}, attempts to cover the entire mean gross income spectrum, including all the highest incomes in the tenth decile ($N_{s}=10$).  Large RMS errors and large instabilities were generated when attempting to fit a lognormal distribution to all ten decile mean incomes. A better fit, however, was obtained for the first 9 deciles, leaving the last decile for a Pareto distribution as seen in Figure \ref{fig:ukFits}b. A Pareto shape factor $\alpha=1.5$ and a maximum income of $x_{max} =\pounds 314K$ were the assumed pairing\citep{druagulescu2001exponential}. (A description of calibrating  the Pareto parameters is given a the end of this section.)

%\begin{figure}
%	\begin{center}		
%	     \resizebox{\columnwidth}{!}{	
%		\includegraphics[angle=-90,scale=1.0]{ukPare5dec.eps}
%		}
%	\end{center}
%	\caption{\small Hybrid Implied Distribution for UK Decile Data with Pareto Tail}
%	\label{fig:paretoUK}
%\end{figure}    
%\begin{figure}
%	\begin{center}		
%	     \resizebox{\columnwidth}{!}{	
%		\includegraphics[angle=-90,scale=1.0]{ukP4dec.eps}
%		}
%	\end{center}
%	\caption{\small Lognormal Fit to UK Mean Decile Data}
%	\label{fig:logUK}
%\end{figure}    
  
% https://www.ons.gov.uk/file?uri=/peoplepopulationandcommunity/personalandhouseholdfinances/incomeandwealth/datasets/effectsoftaxesandbenefitsonhouseholdincomehistoricalpersonleveldatasets/averageincomestaxesandbenefitsofallindividualsretiredandnonretiredbydecilegroup/averageincomestaxesandbenefitsofallindividualsretiredandnonretiredbydecilegroup.xlsx 

%\subsubsection{USA Gross Income by Percentage levels}

\begin{figure}
	\begin{center}		
	     \resizebox{\columnwidth}{!}{	
		\includegraphics[angle=-90,scale=1.0]{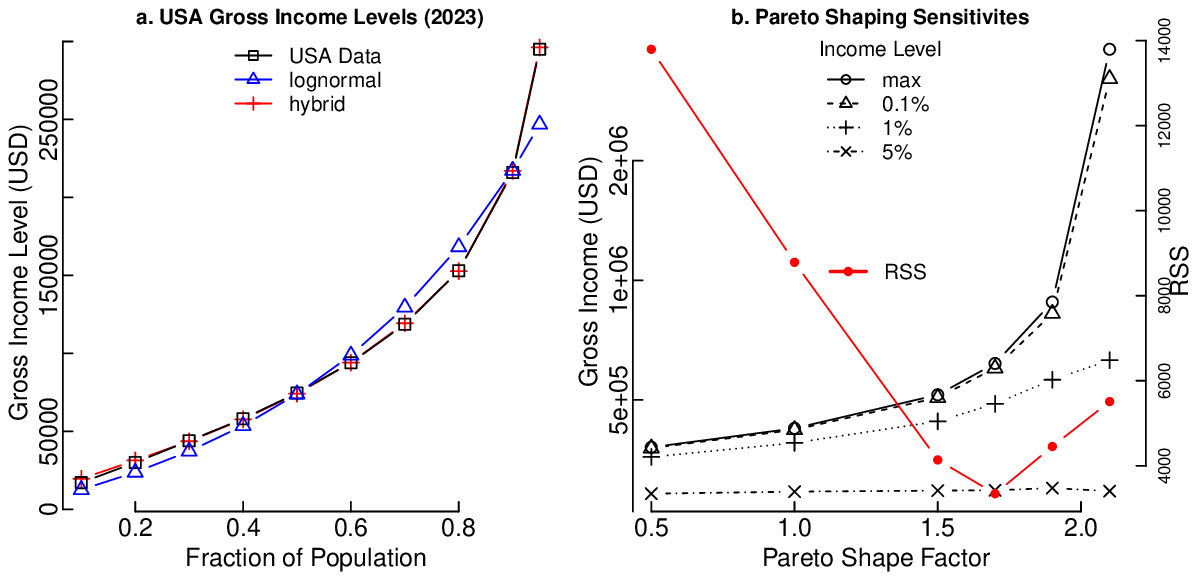}
		}
	\end{center}
	\caption{\small Implied Distributions for USA Data and Pareto Shaping. \scriptsize a) USA data \citep{usa2022} and lognormal and hybrid fits highlighting the benefit of a Pareto tail. b) Pareto fitting error and mean top gross income levels as a function of Pareto shape factor.}
	\label{fig:usdFits}
\end{figure}

From the United States Census for 2023, data levels of gross income that certain percentages of the population fall at or under were acquired. These gross income levels are at every 10\% of the population plus a level at 95\% ($N_{S}=10$) \citep{usa2022}.  The incomes of the remaining top 5\% of the population were left as an exercise for the reader.
RMS errors for lognormal fits were very large and unstable when taken across the full set of income levels. Fitting only the 80\% and below gave a better fit with both low RMS errors and stable model parameters.
%\subsubsection{Pareto Tail Fit}
The last two percentile levels (90\% and 95\%) were fit to the Pareto distribution with the trade off between $\alpha$ and maximum income $x_{max}$.  Figure \ref{fig:usdFits}a compares the USA data with the implied hybrid distribution and the lognormal only distribution in the aggregated format. With the optimal $\alpha=1.7$, Figure \ref{fig:usdFits}b shows the resultant top 1\% mean income as \$488K and a maximum income of (only) \$617K.

%Carpenter\citep{charpentier2022pareto} and Taleb \citep{taleb2015not} discuss this sensitivity in detail.
% \begin{figure}
%	\begin{center}		
%	     \resizebox{\columnwidth}{!}{	
%		\includegraphics[angle=-90,scale=1.0]{usdFit2.eps}
%		}
%	\end{center}
%	\caption{\small Logistic Fit to USA Percentile Data}
%	\label{fig:logUSD}
%\end{figure}    

%https://www.census.gov/library/publications/2023/demo/p60-279.html

%      \begin{figure}
%	\begin{center}		
%	     \resizebox{\columnwidth}{!}{	
%		\includegraphics[angle=-90,scale=1.0]{alphaMaxIg.eps}
%		}
%	\end{center}
%	\caption{\small Pareto Alpha versus Extreme Income}
%	\label{fig:alphas}
%\end{figure}  

%\subsection{ Implied Distributions For Simulated Economies}
Implied income distributions for three emergent distributions were also generated. The three scenarios were no FL, FL with 100\% estate tax (FL no inherit), and FL with no estate tax (FL inherit) ($N_{s}$ for each scenario is given in Table \ref{tab:fits}). The implied hybrid distributions are shown in Figure \ref{fig:modelAndActual}a along with the model's actual income distributions.
Implied distributions for the model's entire populations' income distributions had unusual Pareto shape factors ( $\alpha < 1$) and insignificant improvements in fitting accuracy over the lognormal only fit. 
%Even quadrupling the landscape area and carry capacity did not entice a fatter non-lognormal rich tail as seen by the FL inherit quad attempt in Table \ref{tab:fits}.

%\subsection{Sensitivity of Inequality to Pareto Fit Parameters} \label{section:sensitivityGC}

The process of fitting a discrete Pareto function to the empirical data entails specifying the Pareto shape factor $\alpha$, the Pareto regime threshold $x_{m}$ and the length of the (finite) Pareto tail as represented by the maximum income $x_{max}$. Setting the income level where the lognormal fit has not yet suffered large RSS errors as $x_{m}$, then $\alpha$ and  $x_{max}$ are solved for the minimum RSS error for the remaining empirical data points. For the US case, $x_{m}$ was found to be the 80\% level, leaving the 90\% and 95\% data points for RSS fitting. Figure \ref{fig:usdFits}b displays these values as well as the resultant income levels for top 5\%, 1\% and 0.1\% fractions of the population. For the decile UK data, only the 10 decile datum is available to fit both $\alpha$ and $x_{max}$, requiring an assumed pairing of $\alpha$ and $x_{max}$. 
% The GC metric is a function of population and sample size and has also been shown to be upwardly biased for Pareto tail distributions if the threshold $x_{min}$ is too low \citep{charpentier2022pareto,fontanari2018gini}. 
 %Even GC measurements on entire populations are poor estimators across populations of different sizes or even time-varying populations \citep{fontanari2018gini}. Thus the GC is not reliable metric for this analysis.
%Figure \ref{fig:alphas} provides the sensitivity of the implied tail distribution's optimized $\alpha$ and resultant tail distribution and Gini coefficient based on the estimated maximum $x_{max}$ for the UK and USA data sets described in the following sections.
%If additional data on the tail incomes are available (e.g. top 5\% mean income), this data point may be used instead of the estimated $x_{max}$.

%\subsection{Income Inequality and Redistribution in Real Economies}
\begin{figure}
	\begin{center}		
	     \resizebox{\columnwidth}{!}{	
		\includegraphics[angle=-90,scale=1.0]{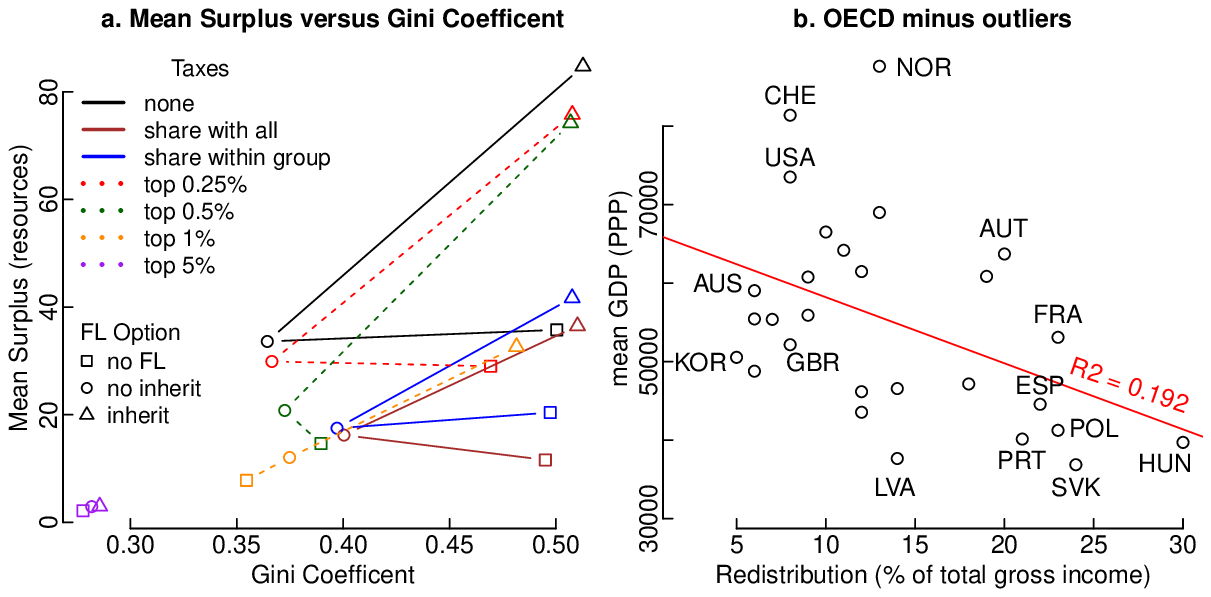}
		}
	\end{center}
	\caption{\small Tax Effects on Mean Wealth and Inequality.  \scriptsize a) FL option effects of income tax rates and estate tax on the simulation's mean surplus and inequality. %For the no income tax configuration, no FL (death only by starvation) and FL with 100\% estate tax both have the same mean surplus with the estate tax option showing a significant reduction in inequality. With FL and inheritance without taxation, mean surplus shows a significant gain with inequality unchanged from no FL. As income taxes increase for the FL configurations, mean surplus decreases with little effect on inequality until mean surplus is drastically reduced to close to zero. 
	b) OECD per capita GDP as purchasing power parity (PPP)  \citep{imfGdp,wbGdp,ciaGdp} versus redistributed per cent of total gross income \citep{causa2017income}. Outliers were LUX, IRL as high GDP and TUR, CHL, and MEX as low GDP.}
	\label{fig:sharing}
\end{figure}
%\citep{drometer2018wealth} 

%/Users/jackcs/gp/figures
\section{Empirical Data and Simulation Comparisons}

Comparisons are made between empirical data and the model's results for the various policies and inheritance scenarios. First the effects on inequality are compared, and then the effects on total income. Finally, the complex interplay between inequality and total wealth on the welfare of the whole population is examined. \hl{}The GC is commonly used by most of the empirical inequality literature despite its well-known shortcomings \citep{fontanari2018gini}. \hl{}One approach to mitigate these problems (used in the cited references) is to report the change in the GC, for example when comparing gross and net income distributions for the same population sample. For the model's distributions, the GC is computed on the whole population with no sampling and with all the populations at the same steady state carry capacity \citep{fontanari2018gini}.
 
% \subsection{Inequality Comparisons}

In the simulations, redistribution effects on total surplus and inequality are given in Figure \ref{fig:sharing}a.  The solid lines represent data points for local-sharing tax and no tax (baseline) scenarios. The dotted lines represent top-down taxing at various levels (by color) of surplus. The shape of the symbols refer to the presence or absence of FL, with or without inheritance (estate tax). Redistribution by income tax (excluding the 5\% bracket) for no FL reduces inequality by up to 30\%, similar to the mean OECD reduction value of 29\% \citep{drometer2018wealth,causa2017income}.  The FL inherit scenario sees reductions no larger than 12\% while the FL no inherit sees no inequality reductions. \hl{}Obviously, the draconian 5\% income tax makes everyone equal but poor with total wealth approaching zero (shown in purple near the origin of Figure \ref{fig:sharing}a).
%\subsection{Human-Scaled Simulation Results}
%The results of these redistribution algorithms on mean surplus and inequality at steady state are presented in Figure \ref{fig:sharing}a. The general \textquote{nose} shape of the runs across FL options (for the no-taxes and the smallest tax bracket populations) reflect two similar inequality measures with inheritance generating populations with twice the mean surplus of no-FL populations.  Increasing local sharing reduces mean surplus while maintaining the \textquote{nose} shape. The peak of the \textquote{nose} shows that the 100\% estate tax population has the same mean wealth as the no-FL population with a dramatic reduction in inequality. The increasing taxing scenarios reduce mean surplus proportionally while pushing only the no-FL populations to greater equality, changing the \textquote{nose"} into a straight line.  The highest tax bracket (color purple in the figure) has decimated the surplus of all the populations resulting in the lowest inequality because everyone is equal and poor.  %\citep{braun2006welfare}

%\begin{figure}
%	\begin{center}		
%	     \resizebox{\columnwidth}{!}{	
%		\includegraphics[angle=-90,scale=1.0]{taxGdp.eps}
%		}
%	\end{center}
%	\caption{\small Taxation effects on GDP (PPP)}
%	\label{fig:gdpTax}
%\end{figure}

%\subsection{Total Income Comparisons}

\textquote{Empirical findings in the literature that growth tends to be inequality neutral} ignores absolute inequality \citep{ravallion2014income,rawls1971atheory}. Overall welfare is dictated by total income as well as its relative distribution. Based on empirical data, some studies have shown increased income taxes reduce gross income (Figure \ref{fig:sharing}b) \citep{trabandt2011laffer,prescott2004americans} while others do not \citep{ravallion2014income,rawls1971atheory}.   The model shows income taxes proportionality reduce total surplus.

A successful redistribution approach for the model turned out to be the estate tax. The addition of FL without inheritance (100\% estate tax) results in a significant reduction in inequality without a reduction in total surplus when compared to no FL. \hl{}The estate taxes collected are not redistributed but lost which represents a worst case utilization of these particular taxes. A strong positive correlation between surplus lost (due to estate tax or no living heirs) and higher total surplus is most likely an effect, not a cause, though cause does find support in the literature: \textcquote{braun2006welfare}{improve welfare by increasing taxes and throwing away tax revenues}. Unfortunately, for comparison purposes, estate taxes in real economies do not generate significant revenue and, therefore, do not impact overall welfare. There is, however, considerable academic literature that strongly suggests estate taxation would be a fair and effective tax to reduce inequality. \citep{caron2012occupy,drometer2018wealth,hoover1927economic,aaron1992reassessing,bird2013death}  

%\subsection{Comparisons of Population Welfare}

Untaxed inheritance scenarios showed a strong increase in total surplus with the inequality measure returning to the baseline no FL value. The question to be addressed is whether the increase in total surplus is sufficient to improve the overall welfare of the middle and poor classes.  Even the combined use of GC and total surplus may not answer this question. To make a determination, the actual surplus distributions need to be examined.

%Figure \ref{fig:modelAndActual} presents emergent distributions as well as hybrid distributions fit to these entire population distributions. 
Figure \ref{fig:modelAndActual}a presents income distributions of entire populations from the model and the hybrid fit to these distributions.
In Figure \ref{fig:modelAndActual}a,  the vertical dotted lines represent the mean surplus for three configurations:  FL with estate tax, no FL, and FL with untaxed inheritance (in order of increasing mean surplus).  The FL with estate tax (FL no inherit) scenario (red) has significantly less inequality than the no FL scenario (orange) for essentially the same mean surplus. The much higher mean surpluses of the FL with inheritance (FL inherit) scenario (blue) highlight the issue. Close examination of the two distributions clearly shows the FL inherit surpluses are all in the rich tail of the distribution and all the agents below the FL inherit mean (the vast majority of the population) are better off with the estate tax policy, even though that tax is a sunk cost, is not redistributed, and has a significantly lower total wealth.
These actual surplus comparisons over the entire populations provide the resolution to estimate the relative welfare each income of the two populations that may be obscured by single point inequality metrics. 
%are not subject to the errors a single point metric like GC has with these differently sized populations. 
Thus the difficult question of the interplay of inequality and total wealth under various redistribution policies can be addressed.
\hl{}Figure \ref{fig:modelAndActual}b \hl{}aggregates these actual distributions and compares them to the US empirical data. In addition to the significant effect of the inheritance tax, the other two simulated distributions show a strong resemblance to the USA empirical data with generally a slightly better population welfare visible in the graph.

\section{Results}
Significant inequality for all lifespan options emerged even with identically capable agents and equal opportunity landscapes. Though no real economy approaches these ideals, inequality as a result of stochastic population dynamics independent of systemic inequalities is an unexpected explanation \citep{petit2010systemic}.

These comparisons between a minimal model of a system and empirical measurements of actual economies highlight their similarities and differences. The former has simple rules, a fixed landscape, easily designed experiments to investigate cause and effect, and clear, quantitative results.  It also provides a full picture of the stochasticity of population and income dynamics. The later has a multitude of economies; all with different policies, assets, and resources; and varying degrees of transparency. The later also provides only one instance of a stochastic process. 
%The measurement of actual inequality is an inexact science with theoretical shortcomings of metrics, data sampling issues especially in the rich tail, and modeling biases. 
%Nonetheless, a number of similarities and differences have been identified. 
Both the model and the empirical data show significant and similar reductions in inequality as measured by percentage change in GC for income-tax-based redistribution. 
%The local-sharing tax redistributions did not affect inequality and were shown to be regressive taxes.  
The most effective redistribution tax was found to be the 100\% estate tax with the poor and middle class better off than either the higher total wealth FL inherit  or the no FL scenarios.
%with the model generating substantial inequality reductions without any total surplus penalties though actual economies do not generate significant revenues from this tax. 
The aggregated income distributions emerging from the simulations show in Figure \ref{fig:modelAndActual}b a striking resemblance to the aggregated empirical data. The model's implied distributions lack significant Pareto tails that were required for good fits on the empirical data.
These generative and implied distributions provided insight into the welfare of the population beyond what simple inequality and aggregated mean income measurements can provide. 
\footnotesize
\bibliographystyle{splncs04}

%\bibliography{compX2.bib} % replace by the name of your .bib file
%\bibliography{feb03_23.bib} % replace by the name of your .bib file
%\bibliography{/Users/jackcs/biblio/feb03_23.small.bib} % replace by the name of your .bib file
%\bibliography{/Users/jackcs/biblio/sep09_23.bib} % replace by the name of your .bib file
%\bibliography{/Users/jackcs/biblio/april_24.bib} % replace by the name of your .bib file
%\bibliography{/home/jackcs/biblio/april_24.bib} % replace by the name of your .bib file
\bibliography{april_24.bib} % replace by the name of your .bib file

%\bibliography{feb03_23.small.bib} % replace by the name of your .bib file

%\begin{thebibliography}{6}
%

%\end{thebibliography}

\end{document}